\documentstyle[12pt]{article}

  \def\pp{{\mathchoice
          {
              \kern 1pt%
              \raise 1pt
              \vbox{\hrule width5pt height0.4pt depth0pt
                    \kern -2pt
                    \hbox{\kern 2.3pt
                          \vrule width0.4pt height6pt depth0pt
                          }
                    \kern -2pt
                    \hrule width5pt height0.4pt depth0pt}%
                    \kern 1pt
           }
            {
              \kern 1pt%
              \raise 1pt
              \vbox{\hrule width4.3pt height0.4pt depth0pt
                    \kern -1.8pt
                    \hbox{\kern 1.95pt
                          \vrule width0.4pt height5.4pt depth0pt
                          }
                    \kern -1.8pt
                    \hrule width4.3pt height0.4pt depth0pt}%
                    \kern 1pt
            }
            {
              \kern 0.5pt%
              \raise 1pt
              \vbox{\hrule width4.0pt height0.3pt depth0pt
                    \kern -1.9pt  
                    \hbox{\kern 1.85pt
                          \vrule width0.3pt height5.7pt depth0pt
                          }
                    \kern -1.9pt
                    \hrule width4.0pt height0.3pt depth0pt}%
                    \kern 0.5pt
            }
            {
              \kern 0.5pt%
              \raise 1pt
              \vbox{\hrule width3.6pt height0.3pt depth0pt
                    \kern -1.5pt
                    \hbox{\kern 1.65pt
                          \vrule width0.3pt height4.5pt depth0pt
                          }
                    \kern -1.5pt
                    \hrule width3.6pt height0.3pt depth0pt}%
                    \kern 0.5pt
            }
        }}

  \def\mm{{\mathchoice
                       {
                             \kern 1pt
               \raise 1pt    \vbox{\hrule width5pt height0.4pt depth0pt
                                  \kern 2pt
                                  \hrule width5pt height0.4pt depth0pt}
                             \kern 1pt}
                       {
                            \kern 1pt
               \raise 1pt \vbox{\hrule width4.3pt height0.4pt depth0pt
                                  \kern 1.8pt
                                  \hrule width4.3pt height0.4pt depth0pt}
                             \kern 1pt}
                       {
                            \kern 0.5pt
               \raise 1pt
                            \vbox{\hrule width4.0pt height0.3pt depth0pt
                                  \kern 1.9pt
                                  \hrule width4.0pt height0.3pt depth0pt}
                            \kern 1pt}
                       {
                           \kern 0.5pt
             \raise 1pt  \vbox{\hrule width3.6pt height0.3pt depth0pt
                                  \kern 1.5pt
                                  \hrule width3.6pt height0.3pt depth0pt}
                           \kern 0.5pt}
                       }}

\catcode`@=11
\def\un#1{\relax\ifmmode\@@underline#1\else
        $\@@underline{\hbox{#1}}$\relax\fi}
\catcode`@=12

\let\du=\du                     

\def\d{\delta}

\def\f{\phi}

\def\h{\eta}

\def\j{\psi}
\def\k{\kappa}
\def\l{\lambda}

\def\o{\omega}
\def\p{\pi}
\def\q{\theta}

\def\x{\xi}

\def\O{\Omega}

\def\ve{\varepsilon}
\def\vf{\varphi}

\def\ck{{\cal K}}

\def\cm{{\cal M}}

\def\bo{{\raise-.5ex\hbox{\large$\Box$}}}               
\def\pa{\partial}                                       
\def\TH{{\raise.2ex\hbox{$\displaystyle \bigodot$}\mskip-4.7mu \llap H \;}}
\def\face{{\raise.2ex\hbox{$\displaystyle \bigodot$}\mskip-2.2mu \llap {$\ddot
        \smile$}}}                                      

\def\sp#1{{}^{#1}}                              
\def\abs#1{\left| #1\right|}                    
\def\leftrightarrowfill{$\mathsurround=0pt \mathord\leftarrow \mkern-6mu
        \cleaders\hbox{$\mkern-2mu \mathord- \mkern-2mu$}\hfill
        \mkern-6mu \mathord\rightarrow$}
\def\dvec#1{\vbox{\ialign{##\crcr
        \leftrightarrowfill\crcr\noalign{\kern-1pt\nointerlineskip}
        $\hfil\displaystyle{#1}\hfil$\crcr}}}           

\def\frac#1#2{{\textstyle{#1\over\vphantom2\smash{\raise.20ex
        \hbox{$\scriptstyle{#2}$}}}}}                   
\def\sfrac#1#2{{\vphantom1\smash{\lower.5ex\hbox{\small$#1$}}\over
        \vphantom1\smash{\raise.4ex\hbox{\small$#2$}}}} 
\def\bfrac#1#2{{\vphantom1\smash{\lower.5ex\hbox{$#1$}}\over
        \vphantom1\smash{\raise.3ex\hbox{$#2$}}}}       
\def\afrac#1#2{{\vphantom1\smash{\lower.5ex\hbox{$#1$}}\over#2}}    

\def\[{\lfloor{\hskip 0.35pt}\!\!\!\lceil}
\def\]{\rfloor{\hskip 0.35pt}\!\!\!\rceil}

\def\du#1#2{_{#1}{}^{#2}}

\def\Tr{{\rm Tr}}

\def\un{\underline}
\def\fracmm#1#2{{{#1}\over{#2}}}

\def\low#1{{\raise -3pt\hbox{${\hskip 0.75pt}\!_{#1}$}}}

\newskip\humongous \humongous=0pt plus 1000pt minus 1000pt
\def\caja{\mathsurround=0pt}
\def\eqalign#1{\,\vcenter{\openup2\jot \caja
        \ialign{\strut \hfil$\displaystyle{##}$&$
        \displaystyle{{}##}$\hfil\crcr#1\crcr}}\,}
\newif\ifdtup

\def\ref#1{$\sp{#1)}$}

\parskip=\medskipamount                 
\thicklines                         

\begin{document}


\thispagestyle{empty}               

\def\border{                                            
        \setlength{\unitlength}{1mm}
        \newcount\xco
        \newcount\yco
        \xco=-24
        \yco=12
        \begin{picture}(140,0)
        \put(-20,11){\tiny Institut f\"ur Theoretische Physik Universit\"at
Hannover~~ Institut f\"ur Theoretische Physik Universit\"at Hannover~~
Institut f\"ur Theoretische Physik Hannover}
        \put(-20,-241.5){\tiny Institut f\"ur Theoretische Physik Universit\"at
Hannover~~ Institut f\"ur Theoretische Physik Universit\"at Hannover~~
Institut f\"ur Theoretische Physik Hannover}
        \end{picture}
        \par\vskip-8mm}

\def\headpic{                                           
        \indent
        \setlength{\unitlength}{.8mm}
        \thinlines
        \par
        \begin{picture}(29,16)
        \put(75,16){\line(1,0){4}}
        \put(80,16){\line(1,0){4}}
      \put(85,16){\line(1,0){4}}
        \put(92,16){\line(1,0){4}}

        \put(85,0){\line(1,0){4}}
        \put(89,8){\line(1,0){3}}
        \put(92,0){\line(1,0){4}}

        \put(85,0){\line(0,1){16}}
        \put(96,0){\line(0,1){16}}
        \put(92,16){\line(1,0){4}}

        \put(85,0){\line(1,0){4}}
        \put(89,8){\line(1,0){3}}
        \put(92,0){\line(1,0){4}}

        \put(85,0){\line(0,1){16}}
        \put(96,0){\line(0,1){16}}
        \put(79,0){\line(0,1){16}}
        \put(80,0){\line(0,1){16}}
        \put(89,0){\line(0,1){16}}
        \put(92,0){\line(0,1){16}}
        \put(79,16){\oval(8,32)[bl]}
        \put(80,16){\oval(8,32)[br]}

        \end{picture}
        \par\vskip-6.5mm
        \thicklines}

\border\headpic {\hbox to\hsize{
\vbox{\noindent DESY 96 -- 259 \hfill December 1996 \\
ITP--UH--25/96 \hfill hep-th/9612170 \\ }}}

\noindent
\vskip1.3cm
\begin{center}

{\Large\bf Mixed (open/closed) N=(2,2) string theory 
\vglue.1in
          as an integrable deformation of self-duality~\footnote{Supported in
part by the `Deutsche Forschungsgemeinschaft' and the `Volkswagen Stiftung'}}\\
\vglue.3in

Sergei V. Ketov \footnote{
On leave of absence from:
High Current Electronics Institute of the Russian Academy of Sciences,
\newline ${~~~~~}$ Siberian Branch, Akademichesky~4, Tomsk 634055, Russia}

{\it Institut f\"ur Theoretische Physik, Universit\"at Hannover}\\
{\it Appelstra\ss{}e 2, 30167 Hannover, Germany}\\
{\sl ketov@itp.uni-hannover.de}
\end{center}

\vglue.3in

\begin{center}
{\Large\bf Abstract}
\end{center}
The exact effective field equations of motion, corresponding to the perturbative
mixed theory of open and closed (2,2) world-sheet supersymmetric strings, are
investigated. It is shown that they are only integrable in the case of an {\it 
abelian} gauge group. The gravitational equations are then stationary with 
respect to the Born-Infeld-type effective action.

\newpage

\section{Introduction}

The {\it closed} (2,2) world-sheet supersymmetric string theory is known 
to have the interpretation of being a theory of {\it self-dual gravity} (SDG) 
\cite{ov}. Similarly, the {\it open} (2,2) string theory can be interpreted as a 
{\it self-dual Yang-Mills} (SDYM) \cite{mar}. Since open strings can `create' 
closed strings which, in their turn, 
can interact with the open ones, there are quantum corrections to the effective 
field equations of the open (2,2) string theory. Because of the  `topological'
nature of the (2,2) string theories, {\it only} 3-point tree string amplitudes 
are non-vanishing and local. As a result, quantum perturbative corrections in the 
mixed theory of open {\it and} closed (2,2) strings are still under control. In
particular, the SDYM equations receive corrections from diagrams with internal
gravitons, so that they become the YM self-duality equations on a K\"ahler
background \cite{mar}. Therefore, they still respect integrability, as expected.
Contrary to the SDYM equations and naive expectations, the effective gravitational
equations of motion in the mixed (open/closed) (2,2) string theory get modified 
in such a way that the resulting `spacetime' is no longer self-dual \cite{mar}. 
Accordingly, the integrability property seems to be lost in the mixed (2,2) 
string theory. I examine the exact effective field equations of motion in the 
mixed theory, and show that the integrability is nevertheless maintained in the 
case of an {\it abelian} gauge group. The effective field theory is {\it not} the
Einstein-Maxwell system describing an interaction of the non-linear graviton with
photon. Instead, it is of the Born-Infeld-type, and is non-linear with respect to
the both fields, gravitational and `electromagnetic'.

In sect.~2 the basic facts about closed and open (2,2) strings separately are
summarized, while the mixed theory is discussed in sect.~3 along the lines of
the Marcus work \cite{mar}. The mixed effective field equations of motion were 
also found first by Marcus \cite{mar}. However, it is the meaning of the 
gravitational equations that remained mysterious, and their possible 
integrability properties were not explored. It is the purpose of this Letter to 
investigate the effective field equations of motion in the mixed (2,2) string 
theory and give the conditions for their integrability.

\section{Basic facts about closed and open (2,2) strings}

The (2,2) strings are strings with {\it two} world-sheet supersymmetries, both 
for the left- and right-moving degrees of freedom.~\footnote{(2,1) and (2,0) 
heterotic strings can also be defined \cite{ovh}. Since the heterotic strings 
have to live \newline ${~~~~~}$ in a $2+1$ or $1+1$ dimensional spacetime where 
self-duality is lost (or hidden, at least), we do \newline ${~~~~~}$ not consider
them here (see, however, ref.~\cite{klm}).} The critical open and closed (2,2) 
strings live in four real dimensions, with the signature $2+2$.  
The physical spectrum consists of a single massless particle, which can be 
assigned in the adjoint of a gauge group $G$ in the open string case.

The only non-vanishing $(2,2)$ string tree scattering amplitudes are 3-point
trees, while all higher $n$-point functions vanish due to kinematical reasons in
$2+2$ dimensions. Tree-level calculations of string amplitudes do not require
the heavy mashinery of BRST quantization~\cite{hann}, or topological methods
\cite{bv}. The vertex operator for a (2,2) closed string particle of momentum $k$
reads in (2,2) world-sheet superspace as
$$ V_{\rm c} =\fracmm{\k}{\p}\exp\left\{ i\left(k\cdot\bar{Z}+\bar{k}\cdot Z
\right)\right\}~,\eqno(2.1)$$
where $\k$ is the $(2,2)$ closed string coupling constant, and 
$Z^i(x,\bar{x},\q,\bar{\q})$ are complex (2,2) chiral superfields.~\footnote{
Throughout the paper, complex coordinates $(x,\bar{x})$ are used for a string
world-sheet, while $(z^i,\bar{z}^{\bar{i}})$ \newline ${~~~~~}$ denote complex 
coordinates of the (2,2) string target space, $i=1,2$.}

When using the $(2,2)$ super-M\"obius invariance of the $(2,2)$ super-Riemann
sphere, it is not difficult to calculate the correlation function of three 
$V_{\rm c}\,$. One finds \cite{ov}
$$ A_{\rm ccc}= \k c_{23}^2~,\quad {\rm where}\quad
c_{23}\equiv \left(k_2\cdot\bar{k}_3-\bar{k}_2\cdot k_3\right)~.\eqno(2.2)$$
One can check that the $A_{\rm ccc}$ is totally symmetric on-shell. and it is
only invariant under the subgroup $U(1,1)\cong SL(2,{\bf R})\otimes U(1)$ of the 
full `Lorentz' group $SO(2,2)\cong SL(2,{\bf R})\otimes SL(2,{\bf R})'$ in $2+2$
dimensions.

Since all higher correlators are supposed to vanish \cite{ov}, the local 3-point
function (2.2) alone determines the {\it exact} effective action \cite{ov},
$$ S_{\rm P} =\int d^{2+2}z\,\left( \fracmm{1}{2}\h^{i\bar{j}}\pa_i\f
\bar{\pa}_{\bar{j}}\f + \fracmm{2\k}{3}\f\pa\bar{\pa}\f\wedge 
\pa\bar{\pa}\f\right)~,\eqno(2.3)$$
which is the {\it Pleba\'nski} action for self-dual gravity (SDG).  
Hence, the massless `scalar' of the closed string theory can be identified 
with a deformation of the K\"ahler potential $K$ of the self-dual 
(=K\"ahler + Ricci-flat) gravity \cite{ov}, where
$$ K=\h_{i\bar{j}}z^i\bar{z}^{\bar{j}}+4\k\f~,\qquad 
\h_{i\bar{j}}=\h^{i\bar{j}}=
\left( \begin{array}{cc} 1 & 0 \\ 0 & -1 \end{array}\right)~.
\eqno(2.4)$$
The (2,2) closed string target space metric is therefore given by 
$$ g_{i\bar{j}}=\pa_i\bar{\pa}_{\bar{j}}K=\h_{i\bar{j}}
+4\k \pa_i\bar{\pa}_{\bar{j}}\f~.\eqno(2.5)$$

Similarly, in the open (2,2) string case, when using the 
$N=(2,2)$ superspace vertex
$$ V_{\rm o} =g\exp\left\{ i\left(k\cdot\bar{Z}+\bar{k}\cdot Z
\right)\right\}~,\eqno(2.6)$$
to be assigned to the boundary of the $(2,2)$ supersymmetric upper-half-plane 
(or $(2,2)$ super-disc) with proper boundary conditions, one finds the 
three-point function \cite{mar}
$$ A_{\rm ooo} =-igc_{23}f^{abc}~,\eqno(2.7)$$
which is essentially a `square root' of $A_{\rm ccc}$, as it should ($f^{abc}$
are structure constants of $G$). The $ A_{\rm ooo}$ can be obtained from the 
effective action \cite{mar}
$$ S_{\rm DNS} = \int d^{2+2}z\,\h^{i\bar{j}}\left( \fracmm{1}{2}\pa_i\vf^a
\bar{\pa}_{\bar{j}}\vf^a - i\fracmm{g}{3}f^{abc}\vf^a\pa_i\vf^b\bar{\pa}_{\bar{j}}
\vf^c\right)+\,\ldots~~.\eqno(2.8)$$
Requiring all the higher-point amplitudes to vanish in the {\it field} theory 
(2.8) determines the additional local $n$-point interactions, $n>3$, which were
denoted by dots in eq.~(2.8). The full action $S_{\rm DNS}$ is known as the 
{\it Donaldson-Nair-Schiff} (DNS) action \cite{dns}. The DNS equation of motion 
is just the {\it Yang} equation \cite{yang} of the SDYM,
$$ \h^{i\bar{j}} \bar{\pa}_{\bar{j}}\left( e^{-2ig\vf}\pa_i e^{2ig\vf}\right)=0~,
\eqno(2.9)$$
where the matrix $\vf$ is Lie algebra-valued, $\vf=\vf^at^a$, and the Lie algebra
generators $t^a$ of $G$ are taken to be anti-hermitian. The DNS action is known 
to be dual (in the field theory sense) to the {\it Leznov-Parkes} (LP) action 
\cite{lez,parkes}, which has only {\it cubic} interaction, and whose equation of 
motion also describes the SDYM.~\footnote{The LP effective action appears from 
the open (2,2) string theory after taking into account the \newline ${~~~~~}$ 
world-sheet instanton corrections~\cite{lecht}. The higher $n$-point functions, 
$n>3$, are known to vanish \newline ${~~~~~}$ in the LP field 
theory \cite{parkes}.} 

\section{Mixed (2,2) string amplitudes and \\ 
         effective field equations of motion}

When open strings join together, they form closed strings. In their turn, the 
closed strings can interact with the open strings. Therefore, the open string 
theory has {\it mixed} (open/closed) amplitudes too. In particular, the 
coupling constants of the closed and open strings are related,
$$ \k \sim \sqrt{\hbar}\,g^2~.\eqno(3.1)$$

The mixed (2,2) string tree amplitudes were also calculated by Marcus 
\cite{mar}. Thus, I can simply `borrow' his results here. The only 
non-vanishing 3-point mixed amplitude is given by
$$ A_{\rm ooc} =\fracmm{\k}{\p} \d^{ab}c_{23}^2 \int^{+\infty}_{-\infty}dx\,
\fracmm{1}{x^2+1}=\k\d^{ab}c_{23}^2~,\eqno(3.2)$$
where the integration over the position $x$ of one of the open string vertices 
goes
along the border of the upper-half-plane (= real line). All higher $n$-point 
mixed amplitudes, $n\geq 4$, are believed to  vanish, like the purely open or 
closed string ones. The additional (mixed) term in the (2,2) open string 
effective field theory action, which has to reproduce the $A_{\rm ooc}$, reads 
as follows \cite{mar}:
$$ S_{\rm M} = \int d^{2+2}z\,\left( 2\k\f\pa\bar{\pa}\vf^a\wedge 
\pa\bar{\pa}\vf^a\right)~.\eqno(3.3)$$
The complete non-abelian effective action can be determined by demanding all 
higher-point amplitudes to vanish in the {\it field} theory describing the mixed 
(2,2) strings, order by order in $n$. Rescaling $\f$ by a factor of $4\k$, and 
$\vf$ by a factor of $g$, one finds \cite{mar}
$$\eqalign{
S_{\rm tot} =&
\fracmm{1}{16\k^2}\int d^{2+2}z\,\left[ \fracmm{1}{2}\h^{i\bar{j}}\pa_i\f
\bar{\pa}_{\bar{j}}\f +\fracmm{1}{6}\f\pa\bar{\pa}\f\wedge \pa\bar{\pa}\f\right]
+\fracmm{1}{g^2}\int d^{2+2}z\,\h^{i\bar{j}}\times  \cr
& \times \left[ -\fracmm{1}{2}\Tr\left(\pa_i\vf\bar{\pa}_{\bar{j}}\vf\right) 
 -\fracmm{2i}{3!}\Tr\left(\bar{\pa}_{\bar{j}}
\vf\[\pa_i\vf,\vf\]\right) +\fracmm{2^2}{4!}\Tr\left(\bar{\pa}_{\bar{j}}\vf
\[\[\pa_i\vf,\vf\],\vf\]\right) +\ldots \right] \cr
& +\fracmm{1}{g^2}\int d^{2+2}z\,\left[ \fracmm{1}{2} \pa\bar{\pa}\f\wedge
\Tr\left(\vf\pa\bar{\pa}\vf\right) +\fracmm{2i}{3!} \pa\bar{\pa}\f\wedge
\Tr\left(\vf\[\pa\vf,\bar{\pa}\vf\]\right)+\ldots\right]~.\cr}\eqno(3.4)$$

Despite of its rather complicated form, the equations of motion resulting from 
that action can be written down in simple geometrical terms \cite{mar}, namely
$$g^{i\bar{j}}(\f)\bar{\pa}_{\bar{j}}\left( e^{-2i\vf}\pa_i e^{2i\vf}\right)=0~,
\eqno(3.5)$$
and
$$-\det g_{i\bar{j}}=+1 +\fracmm{2\k^2}{g^2}\Tr\left(F_{i\bar{j}}F^{i\bar{j}}
\right)~,\eqno(3.6)$$
where $F_{i\bar{j}}$ is the YM field strength of the YM gauge fields
$$A\equiv e^{-i\vf}\pa e^{i\vf}~, \qquad \bar{A}\equiv e^{i\vf}\bar{\pa}
 e^{-i\vf}~, \eqno(3.7)$$
$g_{i\bar{j}}=\h_{i\bar{j}}+\pa_i\bar{\pa}_{\bar{j}}\f$ is a K\"ahler metric,
$g^{i\bar{j}}$ is its inverse, and the indices $(i,\bar{j})$ are raised and 
lowered by using the totally antisymmetric Levi-Civita symbols $\ve^{ij}$,  
$\ve^{\bar{i}\bar{j}}$, and $\ve_{ij}$, $\ve_{\bar{i}\bar{j}}$
$(\ve_{12}=\ve^{12}=1)$.

Eq.~(3.5) is just the Yang equation (of motion) of the DNS action describing the
SDYM on a curved K\"ahler background, as it should have been expected. It is a 
meaning of the gravitational equation (3.6) that is of our interest.
\vglue.2in

\section{Effective field theory of mixed (2,2) strings \\ as an integrable 
deformation of self-duality}

Associated with the K\"ahler metric
$$ ds^2= 2g_{i\bar{j}}dz^id\bar{z}^{\bar{j}}\equiv 2K_{,i\bar{j}}dz^i
d\bar{z}^{\bar{j}}~,\eqno(4.1)$$
is the fundamental (K\"ahler) closed two-form
$$ \O=g_{i\bar{j}}dz^i\wedge d\bar{z}^{\bar{j}}\equiv K_{,i\bar{j}}dz^i\wedge
d\bar{z}^{\bar{j}}~,\eqno(4.2)$$
where $K$ is the (locally defined) K\"ahler potential, and all subscripts after a
comma denote partial differentiations. We regard the complex coordinates 
$(z^i,\bar{z}^{\bar{i}})$ as independent variables, so that our complexified
`spacetime' $\cm$ is locally a direct product of two 2-dimensional complex
manifolds $\cm\cong M_2\otimes\bar{M}_2$, where both  $M_2$ and $\bar{M}_2$ are
endowed with complex structures, i.e. possess closed non-degenerate two-forms
$\o$ and $\bar{\o}$, respectively.~\footnote{The normalization of the holomorphic
two-forms $\o$ and $\bar{\o}$ is fixed by the flat `spacetime' limit \newline
${~~~~~}$ where $\o=dz^1\wedge dz^2$ and $\bar{\o}=d\bar{z}^{\bar{1}}\wedge 
d\bar{z}^{\bar{2}}$.} Hence, the 
effective equations of motion (3.5) and (3.6) in the mixed (2,2) string theory
can be rewritten to the even more geometrical form as \cite{mar}
$$ \O\wedge F=0~,\eqno(4.3)$$
and
$$ \O\wedge \O +\fracmm{4\k^2}{g^2}\Tr (F\wedge F)=2\o\wedge\bar{\o}~,
\eqno(4.4)$$
where $F$ is the YM Lie algebra-valued field strength two-form satisfying
$$ \o\wedge F=\bar{\o}\wedge F=0~.\eqno(4.5)$$

Eqs.~(4.3) and (4.5) are just the {\it self-dual} Yang-Mills equations in the
K\"ahler `spacetime'. They are therefore integrable and their solutions describe 
Yang-Mills instantons (see e.g., the recent paper \cite{japan} and references
therein for some explicit constructions of the solutions). In particular, one
can always locally change the flat SDYM equations of motion into the SDYM 
equations on a curved K\"ahler background by a diffeomorphism transformation
compatible with the K\"ahler structure.

The integrability condition for the gravitational equations of motion in the
complexified `spacetime' is known to be precisely equivalent to the 
(anti)self-duality of the {\it Weyl} curvature tensor \cite{pen}. The famous
twistor construction of Penrose  \cite{pen} transforms the problem of solving
the non-linear partial differential equations of conformally self-dual gravity
into the standard Riemann-Hilbert problem of patching together certain
holomorphic data.

As far as the {\it K\"ahler} spaces are concerned, the self-duality of the Weyl
tensor is precisely equivalent to the vanishing Ricci {\it scalar} curvature
\cite{fla,bpl}, while the Ricci tensor itself is known to be simply related to 
the  K\"ahler metric as
$$ R_{i\bar{j}}=\pa_i\bar{\pa}_{\bar{j}}\log\det(g_{k\bar{k}})~.\eqno(4.6)$$
Eq.~(3.6) or (4.4) therefore yields
$$  R_{i\bar{j}}=\pa_i\bar{\pa}_{\bar{j}}\log\left[ 1+\fracmm{2\k^2}{g^2}\Tr
(F_{i\bar{j}}F^{i\bar{j}})\right]~,\eqno(4.7)$$
and, hence,
$$ R =g^{m\bar{n}}\pa_m\bar{\pa}_{\bar{n}}\log\left[ 1+\fracmm{2\k^2}{g^2}\Tr
(F_{i\bar{j}}F^{i\bar{j}})\right]~,\eqno(4.8)$$
while both do {\it not} vanish on shell. It is also obvious that the `matter'
stress-energy tensor to be equal to the Einstein tensor in accordance with
eqs.~(4.7) and (4.8), does {\it not} vanish too. It is to be compared to the
standard gravitational equations of motion in the case of the Einstein-Yang-Mills 
coupled system to be described by the standard action given by a sum of the
Einstein-Hilbert and the Yang-Mills terms. There, the YM stress-energy tensor
is quadratic with respect to the YM field strength, and it vanishes under the
SDYM condition. In our case, the YM stress-energy tensor is not even polynomial
in the YM field strength, and it has to correspond to a non-polynomial (in $F$)
effective action.

Does it also imply that eq.~(4.4) is not integrable~? First, let us rewrite 
eq.~(3.6) to the form:
$$ \det(g_{i\bar{j}})+\fracmm{2\k^2}{g^2}\Tr\det(F_{i\bar{j}})=-1~,\eqno(4.9)$$
where both determinants are two-dimensional. Given an {\it abelian} field strength
$F$ satisfying the {\it self-duality} condition (4.3) or, equivalently,
$g_{1\bar{1}}F_{2\bar{2}}+g_{2\bar{2}}F_{1\bar{1}}-g_{1\bar{2}}F_{2\bar{1}}-
g_{2\bar{1}}F_{1\bar{2}}=0$, there is a remarkable identity
$$ \det(g) + \fracmm{2\k^2}{g^2}\det(F) = 
\det\left(g+\fracmm{\k\sqrt{2}}{g}F\right)~.\eqno(4.10)$$
In addition, eq.~(3.7) in the abelian case implies
$A=i\pa\vf$, $\bar{A}=-i\bar{\pa}\vf$, and, hence,
$$F=2i\pa\bar{\pa}\vf~.\eqno(4.11)$$
Taken together, they allow us to represent eq.~(4.9) as the Pleba\'nski heavenly
equation
$$ \det\left(\pa\bar{\pa}\ck\right)=-1~,\eqno(4.12)$$
with the {\it complex} potential
$$\ck\equiv K + i\fracmm{2\sqrt{2}\k}{g}\vf~,\eqno(4.13)$$
whose imaginary part is a harmonic function (because of the self-duality of $F$),
and of order $\hbar^{1/2}g$ since eq.~(3.1). Eq.~(4.12) is the consistency 
condition for the linear system
$$\eqalign{
L_{\bar{1}}\j\equiv \left[\bar{\pa}_{\bar{1}}+i\l\bar{B}_{\bar{1}}\right]\j
\equiv 
&  \left[\bar{\pa}_{\bar{1}}+i\l\left(\ck_{,2\bar{1}}\pa_1-\ck_{,1\bar{1}}\pa_2
\right)\right]\j=0~,\cr
L_{\bar{2}}\j\equiv \left[\bar{\pa}_{\bar{2}}+i\l\bar{B}_{\bar{2}}\right]\j
\equiv
&  \left[\bar{\pa}_{\bar{2}}+i\l\left(\ck_{,2\bar{2}}\pa_1-\ck_{,1\bar{2}}\pa_2
\right)\right]\j=0~,\cr}\eqno(4.14)$$
where $\l$ is a complex spectral parameter. The linear equations (4.14) describe 
a fibering of the associated twistor space in the sense of Penrose \cite{pen}.

Hence, one still has the Frobenius integrability, just like that in the usual 
case of the Pleba\'nski heavenly equation with a real K\"ahler potential. The
generalized `metric' to be defined with respect to the complex K\"ahler potential
is not real, and its only use is to make the integrability apparent, while 
the true metric in eq.~(4.1) is real of course. 

When trying to generalize that results to the {\it non-abelian} situation, one 
arrives at an obstruction, since the crucial relation (4.10) is no longer valid. 
If, nevertheless, one wants to impose such a relation, one finds that
$$ F^a\wedge F^b=0~,\quad {\rm when}\quad  a\neq b~.\eqno(4.15)$$
Eq.~(4.15) implies that different directions in the YM group space do not `see' 
each other. Hence, insisting on integrability seems to send us back to the 
abelian case.

It should be noticed that the solutions to the gravitational equations of motion
(4.12) are all stationary with respect to the {\it Born-Infeld}-type effective 
action
$$ S= \int d^{2+2}z\,\sqrt{-\det\left(g_{i\bar{j}}+\fracmm{\k\sqrt{2}}{g}
F_{i\bar{j}}\right)}~.\eqno(4.16)$$
The action $S$ is {\it not} the standard Born-Infeld action \cite{bi}
since the determinant in eq.~(4.16) is two-dimensional, not four-dimensional.
\vglue.2in
 
\section{Conclusion}

Our result is given by the title. To conclude, it is worth mentioning that 
an infinite hierarchy of conservation laws and the infinite number of
symmetries \cite{bpl2} exist as the consequences of Penrose's twistor 
construction when it is formally applied to our `almost self-dual' gravity with 
a complex K\"ahler potential. The underlying symmetry is known to be a loop group
$S^1\to SDiff(2)$ of the {\it area-preserving} (holomorphic) diffeomorphisms (of
a 2-plane), which can be considered as a `large N limit' $(W_{\infty})$ of the
$W_{\rm N}$ symmetries in two-dimensional conformal field theory 
\cite{bakas, park}. The area-preserving holomorphic diffeomorphisms, 
$$ \pa_i\bar{\pa}_{\bar{j}}\ck(z,\bar{z})\to \pa_i\x^k(z)
\pa_k\bar{\pa}_{\bar{k}}\ck(\x,\bar{\x})\bar{\pa}_{\bar{j}}\bar{\x}^{\bar{k}}
(\bar{z})~, \eqno(5.1)$$
leave eqs.~(4.12) and (4.16) invariant, since
$$ \abs{\det(\pa_i\x^k)}=1\eqno(5.2)$$
by their definition.
\vglue.2in

\section*{Acknowledgement} 
I acknowledge discussions with Olaf Lechtenfeld and Alexander Popov.

\end{document}